\documentstyle[11pt,newpasp,twoside,epsf]{article}
\begin{document}
%
%
\title{Ram Pressure and Anomalous Shell Formation in HoII}
\author{M.~Bureau}
\affil{Columbia Astrophysics Laboratory, 550~West 120th~Street,
1027~Pupin Hall, Mail Code~5247, New York, NY~10027, U.S.A.}
\author{C.~Carignan}
\affil{D\'{e}partement de Physique and Observatoire du Mont
M\'{e}gantic, Universit\'{e} de Montr\'{e}al, C.P.~6128,
Succ.~''centre-ville'', Montr\'{e}al, Qu\'{e}bec, Canada, H3C~3J7}
%
%
\begin{abstract}\noindent
  Neutral hydrogen VLA D-array observations of the dwarf irregular
  galaxy HoII, a prototype galaxy for studies of shell formation, are
  presented. The large-scale H$\,${\small I} morphology is reminiscent
  of ram pressure and is unlikely caused by interactions. A case is
  made for intragroup gas in poor and compact groups like the M81
  group, to which HoII belongs. Numerous shortcomings of the supernova
  explosions and stellar winds scenario to create the shells in HoII
  are highlighted, and it is suggested that ram pressure may be able
  to reconcile the observations available.
\end{abstract}
%
%
\section{Introduction}
HoII is a dwarf irregular galaxy on the outskirts of the M81 group, at
a distance of 3.2~Mpc ($M_B$=$-17.0$~mag). It is one of the first
galaxies outside the Local Group where the effects of sequential star
formation on the interstellar medium (ISM) were investigated. Puche et
al.\ (1992; hereafter PWBR92) present high-resolution
multi-configuration VLA H$\,${\small I} observations, revealing a
complex pattern of interconnected shells and holes. They argue for
self-propagating star formation, whereby supernova explosions (SNe)
and stellar winds shape the ISM. While we do not wish to challenge the
general relevance of such scenarios here, we will highlight numerous
problems they face in the particular case of HoII. Some have been
noted before, but others are mentioned for the first time.

We reanalyzed PWBR92's H$\,${\small I} observations of HoII, keeping
only the D-array data. We produced a continuum-subtracted
naturally-weighted cube cleaned to a depth of 1$\sigma$
(2.75~mJy~beam$^{-1}$) and associated moment maps. The total
H$\,${\small I} map is shown in Figure~1 superposed on an optical
image. A previously undetected, large but faint component extends over
the entire northwest half of the galaxy, encompassing the H$\,${\small
I} cloud detected by PWBR92. The H$\,${\small I} on the southeast side
is compressed, giving rise to a striking NW-SE asymmetry, suggesting
that HoII is affected by ram pressure from an intragroup medium
(IGM). The velocity field shows a clear differentially rotating disk
pattern in the inner 7--8\arcmin, but the kinematics at larger radii
is rather disturbed. The total H$\,${\small I} flux
$F_{\mbox{\scriptsize HI}}=267$~Jy~km~s$^{-1}$, corresponding to
$6.44\times10^8$~M$_{\sun}$.
%
%
\begin{figure}[t]
\plotone{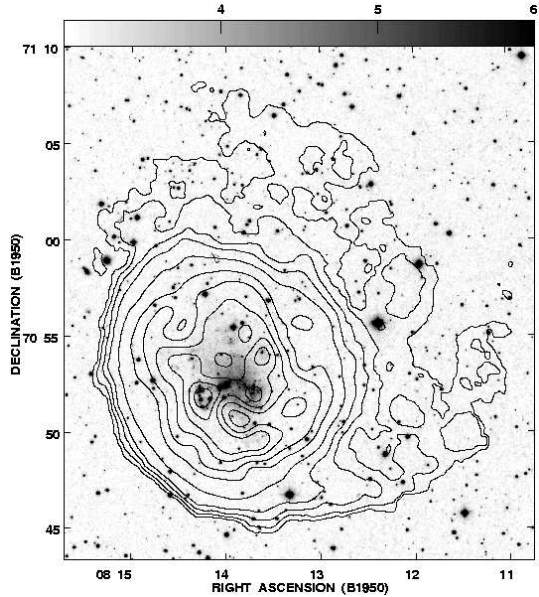}
\caption[]{Total H$\,${\small I} map of HoII from the VLA D-array
data, superposed on a DSS image. Contours are 0.005, 0.015, 0.03,
0.05, 0.10, 0.20, 0.30, 0.45, 0.60, 0.75, and 0.90 times the peak flux
of 8.5~Jy~beam$^{-1}$~km~s$^{-1}$ ($2.10\times10^{21}$~atoms~cm$^{-2}$
or 16.8~M$_{\sun}$~pc$^{-2}$). The beam is
$66\farcs7\times66\farcs7$. The NW-SE asymmetry is obvious.}
\end{figure}
%
%
\section{The Environment of HoII}
The H$\,${\small I} morphology in Figure~1 is reminiscent of ram
pressure but could also be due to interactions. HoII is 475~kpc in
projection from the M81 group center. At the (deprojected) group
velocity dispersion of 190~km~s$^{-1}$ (Huchra \& Geller 1982), it
would take HoII a fifth of a Hubble time to reach the center of the
group. HoII appears to be part of a subsystem of three dwarf irregular
galaxies, with Kar52 (M81dwA) and UGC4483. If HoII is interacting, it
must be with one of these. Kar52 and UGC4483 are much smaller and
fainter than HoII and have irregular optical morphologies, but neither
shows obvious signs of interactions (Bremnes, Binggeli, \& Prugniel
1998). In H$\,${\small I}, Kar52 displays an incomplete lumpy ring
with little rotation (Sargent, Sancisi, \& Lo 1983), and UGC4483 shows
a peaked distribution with a faint envelope extending NW-SE (van Zee,
Skillman, \& Salzer 1998). The distance to HoII is identical to that
of UGC4483 but also to that of NGC2403 and DDO44 to the SW, suggesting
that HoII belongs to the NGC2403 subgroup, along with Kar52, UGC4483,
NGC2365, and DDO44. Karachentsev et al.\ (2000) suggest that the
NGC2403 subgroup is moving towards the M81 group at a velocity of
110-160~km~s$^{-1}$. The environment of HoII thus does not support
interactions as a likely mechanism to shape its large-scale structure,
but rather suggests that it could have a large velocity relative to a
putative IGM. Ram pressure must therefore be considered seriously to
explain its H$\,${\small I} morphology. H$\,${\small I} observations
of the entire region around HoII, Kar52, and UGC4483 should help
clarify this issue and are underway.
%
%
\section{IGM and X-Rays in Small Groups}
The condition for ram pressure stripping can be written as
\begin{equation}
\rho_{\mbox{\tiny IGM}}v^2>2\pi G\Sigma_{tot}\Sigma_{g}
\end{equation}
(Gunn \& Gott 1972), where $\rho_{\mbox{\tiny IGM}}$ is the IGM
density, $v$ the relative velocity of the galaxy with respect to the
IGM, and $\Sigma_{tot}$ and $\Sigma_{g}$ the total and ISM surface
densities, respectively. Taking
$v\approx\sqrt{3}\sigma\approx190$~km~s$^{-1}$ and $\Sigma_{tot}$ and
$\Sigma_{g}$ (corrected for other gaseous species) at the first
significantly disturbed contour in Figure~1, we derive a critical
density for stripping $\rho_{\mbox{\tiny
IGM}}\ga4.0\times10^{-6}$~atoms~cm$^{-3}$.

A virial mass of $1.13\times10^{12}$~M$_{\sun}$ is derived from the
main members of the M81 group (Huchra \& Geller 1982). Spreading 1\%
of this mass in a sphere just enclosing HoII, we obtain a mean density
of $1.4\times10^{-6}$~atoms~cm$^{-3}$, three times less than that
required for stripping. This number provides a benchmark with which to
compare more realistic calculations. The IGM will be more concentrated
and clumpy, and the encounter may not be exactly ``face-on'', but
since the group velocity dispersion is based only on the largest
galaxies, it is probably underestimated, and the group is in any case
unlikely to be virialized at the distance of HoII, making its velocity
highly unconstrained. If HoII is bound to the M81 group, then the
virial mass adopted is also severely underestimated. All these factors
can easily bring the required and derived IGM densities in
agreement. Typical parameters for poor groups are
$R_{vir}\sim0.5h^{-1}$~Mpc and
$M_{vir}\sim0.5-1\times10^{14}h^{-1}$~M$_{\sun}$ (Zabludoff \&
Mulchaey 1998), of which only 10--20\% is in individual galaxies,
leading to a mean density for the remaining matter of
$\sim4\times10^{-3}$~atoms~cm$^{-3}$ within $R_{vir}$. If only 0.1\%
of this is ordinary interacting baryonic matter, then its density is
sufficient to strip the outer ISM of galaxies like HoII. This is
promising since on scales of the virial radius, the dominant baryonic
component in groups is the IGM. Zabludoff \& Mulchaey (1998) report
X-ray gas masses of $1\times10^{12}h^{-5/2}$~M$_{\sun}$ for their
groups, leading to mean densities for the hot gas of
$\sim9\times10^{-5}$~atoms~cm$^{-3}$ within $R_{vir}$.

The total X-ray luminosity of groups does not correlate with the
number of galaxies or optical luminosity, but it does with the
velocity dispersion and gas temperature. A fit to cluster and compact
groups yields, for $\sigma=110\pm10$~km~s$^{-1}$,
$L_X=10^{39.6\pm1.7}$~erg~s$^{-1}$ and $T_{\mbox{\tiny
IGM}}=10^{-0.91\pm0.13}$~keV (Ponman et al.\ 1996). The correlation
for loose groups alone yields $L_X=10^{40.5\pm3.6}$~erg~s$^{-1}$ and
$T_{\mbox{\tiny IGM}}=10^{-0.48\pm0.10}$~keV (Helsdon \& Ponman
2000). The large errors are probably related to the wind injection
histories of the groups, which in turn lead to shallow surface
brightness profiles. There are also indications that the groups and
clusters correlations are different, so both $L_X$ and $T_{\mbox{\tiny
IGM}}$ are probably underestimates, and there can be a large amount of
hot gas in M81-like groups.

Following Cowie \& McKee (1977), the evaporation timescale for a
typical cloud ($n\approx1$~cm$^{-3}$, $R\approx10$~pc) embedded in an
IGM at the aforementioned temperatures is $6.2\times10^5$ to
$2.8\times10^7$~yr. The H$\,${\small I} ``tail'' in HoII extends over
$7-8\arcmin$ in the radial direction. At 190~km~s$^{-1}$, it takes
HoII $3.6\times10^7$~yr to cross that distance. Given the strong
dependence of the evaporation on the assumed properties of the clouds
and IGM, the timescales calculated seem consistent with
observations. In the conditions of interest here, cooling and viscous
stripping (Nulsen 1982) are negligible compared to evaporation.
%
%
\section{The Creation of Shells and Supershells}
PWBR92 studied over 50 H$\,${\small I} holes in HoII and argued for
their formation through SNe and stellar winds. However, H$\alpha$ and
far-UV emission do not preferentially fill small holes or trace the
edges of large ones. The shells are also devoid of hot gas, and X-ray
emission is not preferentially associated with H$\,${\small II}
regions or H$\,${\small I} holes (Kerp \& Walter 2001). The SN rates
derived from radio continuum observations and the H$\,${\small I}
shells agree (Tongue \& Westpfahl 1995), but the energy is deposited
in the central regions of HoII only, hardly helping to explain the
formation of the outer shells. Furthermore, when useful limits are
derived, most stellar clusters expected from massive star formation
are not seen (Rhode et al.\ 1999). Multi-wavelength observations thus
pose a challenge to SNe and stellar winds scenarios for the formation
of the shells in HoII, particularly in the outer parts of the disk,
where no star formation is expected or taking place.

Rhode et al.\ (1999) discuss other mechanisms for the formation of the
shells. SN are most likely not spherically expanding in a uniform ISM,
as assumed, the initial mass function could be very top-heavy, and
gamma-ray bursts could also create holes, but all these mechanisms
still require massive star formation in the outer parts of the disk. A
fractal H$\,${\small I}, overpressured H$\,${\small II} regions,
external ionization sources, and/or high-velocity clouds can bypass
this requirement.

We suggest here that ram pressure provides another solution to the
shell formation problem in HoII. Ram pressure can create holes in an
H$\,${\small I} disk where local minima in the surface density
exist. It provides a mechanism to enlarge pre-existing holes, created
by SNe or otherwise, and can explain the large energy requirements (or
lack of observational signature) from SNe and stellar winds. Of
course, ram pressure-driven shell formation should be properly modeled
before making further claims. It should be easily distinguished from
internal, pressure-driven events, as the shells will have a
``bullet-hole'' geometry like that caused by high-velocity clouds. In
HoII's case, a direct proof of a dense IGM must also be found before
any ram pressure model can be taken seriously.
%
%

%

\begin{references}
\vspace*{-1.75mm}
{\small
\reference Bremnes, T., Binggeli, B., \& Prugniel, P.\ 1998, \aaps, 129, 313
\reference Cowie, L.\ L., \& McKee, C.\ F.\ 1977, \apj, 211, 135
\reference Gunn, J.\ E., \& Gott, J.\ R.\ III 1972, \apj, 176, 1
\reference Helsdon, S.\ F., \& Ponman, T.\ J.\ 2000, \mnras, 315, 356
\reference Huchra, J.\ P., \& Geller, M.\ J.\ 1982, \apj, 257, 423
\reference Karachentsev, I.\ D., et al.\ 2000, \aap, 363, 117
\reference Kerp, J., \& Walter, F.\ 2001, in {\em X-ray Astronomy 2000}, ed.\ R.\ Giacconi, in press.
\reference Nulsen, P.\ E.\ J.\ 1982, \mnras, 198, 1007
\reference Ponman, T.\ J., Bourner, P.\ D.\ J., Ebeling, H., \& B\"{o}hringer, H.\ 1996, \mnras, 283, 690
\reference Puche, D., Westpfahl, D., Brinks, E., \& Roy, J.-R.\ 1992,
\aj, 103, 1841 (PWBR92)
\reference Rhode, K.\ L., Salzer, J.\ J., Westpfahl, D.\ J., \& Radice, L.\ A.\ 1999, \aj, 118, 323
\reference Sargent, W.\ L.\ W., Sancisi, R., \& Lo, K.\ Y.\ 1983, \apj, 265, 711
\reference Tongue, T.\ D., \& Westpfahl, D.\ J.\ 1995, \aj, 109, 2462
\reference Zabludoff, A.\ I., \& Mulchaey, J.\ S.\ 1998, \apj, 496, 39
\reference van Zee, L., Skillman, E.\ D., \& Salzer, J.\ J.\ 1998, \aj, 116, 1186

}
\end{references}
\end{document}